# Spectrum Sensing: Enhanced Energy Detection Technique Based on Noise Measurement

Youness Arjoune[1], Zakaria El Mrabet[1], Hassan El Ghazi[2], and Ahmed Tamtaoui[2]

[1]Electrical Engineering Department
University of North Dakota
Grand Forks, USA

[2]National Institute of Posts and Telecommunications
Rabat, Morocco

*Abstract*—Spectrum sensing enables cognitive radio systems to detect unused portions of the radio spectrum and then use them while avoiding interferences to the primary users. Energy detection is one of the most used techniques for spectrum sensing because it does not require any prior information about the characteristics of the primary user signal. However, this technique does not distinguish between the signal and the noise. It has a low performance at low SNR, and the selection of the threshold becomes an issue because the noise is uncertain. The detection performance of this technique can be further improved using a dynamic selection of the sensing threshold. In this work, we investigate a dynamic selection of this threshold by measuring the power of noise present in the received signal using a blind technique. The proposed model was implemented and tested using GNU Radio software and USRP units. Our results show that the dynamic selection of the threshold based on measuring the noise level present in the received signal during the detection process increases the probability of detection and decreases the probability of false alarm compared to the ones of energy detection with a static threshold.

I. INTRODUCTION

Cognitive radio technology has been proposed as a smart solution to enhance the access to the radio spectrum and solve the problem of its scarcity [1, 2]. Cognitive radio systems enable the detection of unused spectrum holes, and allow secondary users to use them while avoiding interferences to the primary user (PU). To enable dynamic spectrum access, these systems perform spectrum sensing to decide on the presence or absence of the primary user [2, 3]. Several sensing techniques have been proposed to sense the radio spectrum including energy detection [4-8], autocorrelation [9-12], and matched filter based sensing [13,14].

Energy detection [4-8] computes the energy of the received N samples as the squared magnitude of the FFT averaged over these N samples and compares it to a threshold to obtain the sensing decision. If this energy is higher than this threshold, then the primary user is deemed to be present; otherwise, the primary user signal is considered absent. This technique is simple as it does not require any prior information about the primary user signal, which makes it much more straightforward than matched filter and autocorrelation sensing technique [5]. However, the performance of this technique is highly dependent on the noise which is random [6]; thus, using a static threshold degrades the performance of energy detection[7, 8]. A prior knowledge of the noise power or a reliable estimate of its level is thus necessary to enhance its detection performance.

The autocorrelation-based sensing technique [9-12] computes the correlation function of the received N samples with the time-shifted version of these N samples at lag zero and at lag one. If the value of this function is higher than a certain threshold, then the primary user signal is considered present; otherwise, it is considered absent. Because the noise is uncorrelated, the autocorrelation-based sensing technique can distinguish between the signal and noise. The accuracy of this sensing technique depends on the number of samples and the threshold selection. The main drawback of this technique is that it requires a large number of samples to achieve a good performance, and thus, it increases the sensing time, which is not practical for cognitive radio systems that are expected to be rapid and efficient [8].

Matched filter detection [13, 14] is a technique that matches the received samples with some pre-collected and saved pilots of the same primary user signal stream. The received samples are convoluted with the saved pilots then averaged over N samples to compute the decision statistic, which is then compared to a threshold to obtain the sensing decision. If the result of this convolution is higher than a threshold, the primary user signal is considered present; otherwise, it is deemed to be absent. This technique provides better detection at low signal to noise ratio, and it is optimal in the sense that it needs only a few samples to achieve high detection probability in a short sensing time. However, matched filter detection requires the prior knowledge of some of the primary user's signal characteristics. This knowledge is often unavailable, which makes this technique unpractical [8].

According to [5, 6, 9], The performance of these three sensing techniques depends on the number of samples and the sensing threshold. Taking more samples can enhance the detection performance of these techniques up to a certain value of SNR, after which further increase in the number of samples does not improve their detection performance. Increasing the number of samples can also increase the sensing time and in some cases, for instance, the wideband sensing, it is impractical to increase the number of sample when researcher are demanding to use compressive sensing to minimize the number of samples [15-16]. This detection performance can also be enhanced by using a dynamic threshold adapted to the level of noise present in the received signal. In this work, we consider energy detection to show that measuring the level of noise and using it to compute the sensing threshold can further enhance its detection performance.

Last decades has seen several works investigating the improvement of the detection performance of energy detection using a dynamic selection of the threshold [17-22]. For instance, the authors of [17] have proposed a dynamic selection of the threshold using Discrete Fourier Transform Filter Bank method to minimize the spectrum sensing error in a noisy environment. This technique uses the Gradient-based updates to set a new value of the sensing threshold. The authors of [18] proposed an adaptive threshold detection algorithm based on an image binarization technique. This technique dynamically estimates the threshold based on previous iteration decision statistics and other critical parameters such as SNR, the number of samples, and the targeted probabilities of detection and false alarm. In [19], the authors proposed an adaptive threshold that consists of two control parameters to adapt the requirements utilizing targeted probability of detection and false alarm. They determined the threshold based on two methods: Constant False Alarm Rate method which consists of fixing a target probability of false alarm and Constant Detection Rate method which sets the target probability of detection. Then, the threshold that Minimizes spectrum sensing error (MSSE) is selected. The authors of [20, 23] proposed a double-threshold algorithm that consists of using two thresholds $\lambda_1$ and $\lambda_2$ where $\lambda_1 < \lambda_2$ instead of using a single-threshold. If the energy of the samples is smaller than $\lambda_1$ then the spectrum is free and if the energy of samples is higher than $\lambda_2$ then the spectrum is occupied; otherwise, the secondary user is not sure about the presence or absence of the primary user. However, the authors did not explain how these two thresholds were selected. The authors of [22] addressed the process of the threshold selection by using Constant the False Alarm Rate method. This method consists of making a bound on the probability of false alarm and then maximizing the detection probability. The selection of this threshold is dynamically adapted to noise level present in the received signal. However, the authors did not explain how the noise was estimated, and they used directly the value of noise added to the signal generated in Matlab to validate their approach.

All these previously mentioned papers proposed a dynamic selection of the threshold using different algorithms that take into consideration several parameters such as the noise present in the received signal and the target probability of false alarm. They validated their approaches using simulation. However, in a real-world scenario, several system parameters that are assumed constant in simulations may vary over time, which can result in the wrong sensing decision. Thus, the validation of the proposed models through only simulations using Matlab is not enough.

In this work, we investigate the use of dynamic threshold based on measuring the noise level present in the received signal to enhance the probability of detection and decrease the probability of false alarm. The approach measures the noise using a technique based on the eigenvalues of the sample covariance matrix of the received signal. This technique calculates the eigenvalues, then, uses the Minimum Description Length criterion to split the eigenvalues corresponding to the signal and the ones corresponding to the noise [24- 27]. This technique is considered as blind estimation technique because the power of the signal and the power of the noise are unknown and these parameters are estimated from the received signal.

To set the structure of this paper, the rest of this paper is organized as follows. The second section describes the mathematical model of the proposed approach as well as the experimental setup using the GNU Radio software and USRP units. The third section gives and discusses the results. Finally, the conclusions and future research directions are drawn in the conclusion section.

## II. METHODOLOGY

### A. Mathematical Model

Spectrum sensing is one of the most important processes in the cycle of cognitive radio. It aims essentially to decide between two states: primary user signal is absent, denoted by $\mathcal{H}_0$, or primary user signal is present, denoted by $\mathcal{H}_1$. These two states can be modeled as follows:

$$\mathcal{H}_0: y(n) = w(n) \quad (1)$$
$$\mathcal{H}_1: y(n) = x(n) + w(n) \quad (2)$$

where $y(n)$ denotes the received signal, $x(n)$ denotes the transmitted signal, and $w(n)$ denotes the noise affecting the transmitted signal.

In the following, we describe the concept of energy detection and the algorithm used for noise estimation.

#### 1) Energy Detection

As shown in the Fig. 1, energy detection computes the energy of the received N samples as the squared magnitude of the Fast Fourier Transform (FFT) of these samples averaged over N samples [4-8], using the following formula:

$$TED = \sum_{n=1}^{N}(Y[n])^2 \quad (3)$$

This energy $TED$ is then compared to a pre-defined threshold $\lambda_D$ to obtain the sensing decision as follows:

$$TED < \lambda_D : \text{PU signal absent}$$
$$TED > \lambda_D : \text{PU signal present}$$

The detection performance of the algorithm can be evaluated through the probability of detection $P_D$ and the probability of false alarm $P_{FA}$. The probability of detection refers to the numbers of correct detections (PU is present) over the total number of sensing trials while the probability of false alarm refers to the number of times that the PU is falsely detected over the total number of trials. These probabilities are given as:

$$P_d = \Pr(TED > \lambda_D; H_1) \quad (4)$$
$$P_{FA} = \Pr(TED > \lambda_D; H_0) \quad (5)$$

Where $TED$ corresponds to the energy of N samples given by $Eq.\,3$ and $\lambda_D$ is the sensing threshold.

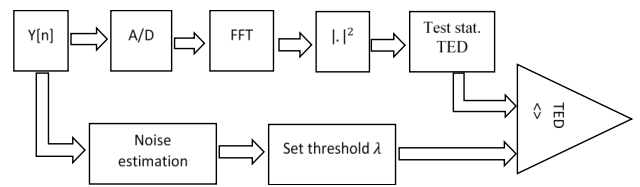

Fig. 1. Energy detection

According to [15], the probabilities of detection $P_d$ and false alarm $P_{fa}$ are given by:

$$P_d = Q\left(\frac{\lambda - N(\sigma_W^2 + \sigma_x^2)}{(\sigma_W^2 + \sigma_x^2)\sqrt{2N}}\right) \quad (6)$$

$$P_{fa} = Q\left(\frac{\lambda_D - N\sigma_W^2}{\sigma_W^2\sqrt{2N}}\right) \quad (7)$$

Where Q-function $Q(x) = \frac{1}{2\pi}\int_x^\infty \exp\left(\frac{-u^2}{2}\right) du$, $\sigma_w$ and $\sigma_s$ are the standard deviation of the noise and the PU signal, respectively, and N is the number of samples.

In [19], the authors gives the formula of the detection-threshold $\lambda_D$, which is obtained from Eq. 7 and given by:

$$\lambda_D = \sigma_W^2(Q^{-1}(P_{fa})\sqrt{2N} + N)) \quad (8)$$

Where $Q^{-1}(.)$ is the inverse of the Q-function.

This threshold depends on the noise variance, the number of samples, and the target probability of false alarm. Thus, a prior knowledge on the level of noise affecting the received signal is required to set an accurate threshold.

*2) Noise Estimation*

The noise is estimated based on the eigenvalues of the sample covariance matrix of the received signal [24-27]. In the following, this technique is described.

Consider the received signal, $y$, which can be expressed as a $N \times L$ matrix:

$$y = \begin{pmatrix} y_{1,1} & \cdots & y_{1,N} \\ \vdots & \ddots & \vdots \\ y_{L,1} & \cdots & y_{L,N} \end{pmatrix} \quad (9)$$

Where $y_{i,j}$ denotes the vector of the received signal samples.

The noise and the signal's samples are assumed to be independent, and also the noise is assumed to be additive white Gaussian noise with mean 0 and variance $\sigma_W^2$. Therefore, Eq. 1 and Eq. 2 can be rewritten as:

$$\mathcal{H}_0: y_i(n) = w_i(n) \quad (10)$$

$$\mathcal{H}_1: y_i(n) = x_i(n) + w_i(n) \quad (11)$$

Where $y_i$ denotes the received signal component, $x_i$ denotes the transmitted component, and $w_i$ denotes the noise component. Consider an observation bandwidth $B$, the transmitted signal with an occupied bandwidth $b$ in the sample covariance matrix eigenvalues domain, and $K \leq L$, the $\frac{K}{L}$ fraction of the whole observation bandwidth is occupied by the transmitted signal, and the rest of the bandwidth is the occupied by the noise. When $L, N \to \infty$, the statistical covariance matrices of the noise, of the transmitted samples, and of the received samples can be defined as:

$$\Sigma_W = E\{w(n)w^H(n)\} = \sigma_W^2. I_L\ ; -\infty < n < +\infty \quad (12)$$

$$\Sigma_x = E\{x(n)x^H(n)\} \quad (13)$$

$$\Sigma_y = E\{y(n)y^H(n)\} \quad (14)$$

Where $\Sigma_W$ denotes the noise statistical covariance matrix, $\Sigma_x$ denotes the transmitted signal statistical covariance matrix, $\Sigma_y$ denotes the statistical covariance matrix of the received signal, $(.)^H$ denotes complex conjugate transpose operator, $\sigma_W^2$ denotes the noise variance, and $I_L$ denotes the L-order identity matrix. Since the signal and the noise are independent, $\Sigma_y$ can be re-written as the sum of $\Sigma_x$ and $\Sigma_w$:

$$\Sigma_y = \Sigma_x + \Sigma_W = \Sigma_x + \sigma_W^2. I_L \quad (15)$$

Given the eigenvalues $\lambda_y$ of $\Sigma_y$ and $\lambda_x$ of $\Sigma_s$ in a descending order, we obtain the following equations:

$$\lambda_{yi} = \lambda_{yi} + \sigma_W^2, \forall i = 1,2 \ldots K \quad (16)$$

$$\lambda_{yi} = \sigma_W^2, \forall i = K+1, K+2, \ldots L \quad (17)$$

Where $\lambda$ denotes the group of eigenvalues and the statistical covariance matrix eigenvalues are equal to signal components power. The estimate of the received statistical covariance matrix $\widehat{\Sigma}_y$ can be computed instead of statistical covariance matrix as there exists a finite number of samples. The sample covariance matrix of the received signal is thus given by:

$$\widehat{\Sigma}_y = \frac{1}{N} yy^H \quad (18)$$

According to the authors of [24-27], the eigenvalues of the samples covariance matrix deviate from the signal power components and follow Marcenko Pastur density, which is dependent on the value of the fraction $\frac{L}{N}$. The value of K is then estimated using the Minimum Descriptive Length criterion. The estimated value of K, denoted as $\widehat{K}$, is given by:

$$K = arg\min_K\left(-(L-K)N \log\frac{\varphi(K)}{\theta(K)} + \frac{1}{2}K(2L - K)\log N\right); 0 \leq K \leq L-1 \quad (19)$$

Where $\varphi(K)$ and $\theta(K)$ are given by:

$$\varphi(K) = \prod_{i=K+1}^L \lambda_i^{\frac{1}{L-K}} \quad (20)$$

$$\theta(K) = \frac{1}{L-K}\sum_{i=M+1}^L \lambda_i \quad (21)$$

Where L is the number of eigenvalues, N is the number of samples and $\lambda_i$ is the set of eigenvalues. After estimating the value of $K$, and according to the same authors of [24-27], the signal group of eigenvalues is determined as $\lambda_1 \ldots \lambda_K$ and the noise group eigenvalues as $\lambda_{K+1} \ldots \lambda_L$.

To compute the noise variance $\sigma_W^2$, two values $\sigma_{W1}^2$ and $\sigma_{W2}^2$ are calculated as follows:

$$\sigma_{W1}^2 = \frac{\lambda_L}{(1-\sqrt{p})^2} \quad (22)$$

$$\sigma_{W2}^2 = \frac{\lambda_{\widehat{K}+1}}{(1+p)^2} \quad (23)$$

Consider M linearly spaced values in the range $[\sigma_{W1}^2, \sigma_{W2}^2]$ are denoted as $\pi_m$, where $1 \leq m \leq M$. The Marcenko Pastuer density of parameters $p, \sigma_w$, is given by:

$$mp(p, \sigma_w) = dF^W(Z)$$

$$= \frac{\sqrt{(z - \sigma_W^2(1-\sqrt{p})^2)(\sigma_W^2(1+\sqrt{p})^2 - z)}}{2\pi\sigma_W^2 zp} dz \quad (24)$$

Where

$$\sigma_W^2(1-\sqrt{p})^2 \leq z \leq \sigma_W^2(1+\sqrt{p})^2 \quad (25)$$

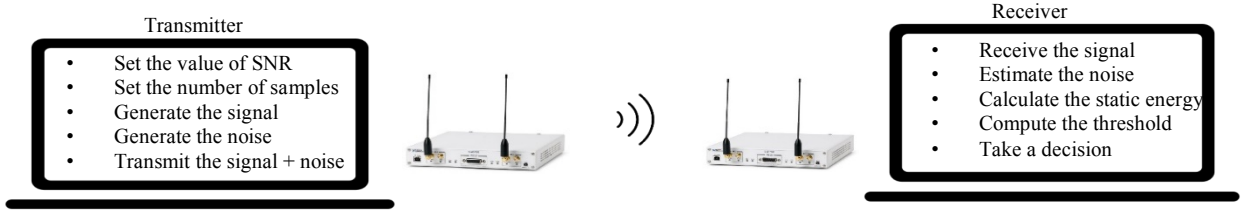

Fig. 2. Experimental setup

Using Eq. 24, we can calculate K Marcenko Pastur densities of the parameters $(1-\hat{\beta})p$ and $\pi_m$ where $\hat{\beta} = \frac{\hat{K}}{L}$ and the Empirical Distribution function (EF) of the noise group eigenvalues is given by,

$$EF = F_n(t) = \frac{number\ of\ samples\ values \leq t}{n} \quad (26)$$

Where n is the total number of sample values in the noise eigenvalues. The noise eigenvalues empirical distribution is then compared with the Marcenko Pastur densities and a goodness of fitting is used to pick the best estimate of $\pi_m$ in order to estimate the value of $\sigma_W^2$. $G(\pi_m)$ denotes the goodness of fitting, and it is given by:

$$G(\pi_m) = \|EF - mp(((1-\hat{\beta})p, \pi_m))\|_2 \quad (27)$$

The estimate of the noise variance, $\widehat{\sigma_W^2}$, is thus given by:

$$\widehat{\sigma_W^2} = \min_{\pi_m}(G(\pi_m)) \quad (28)$$

Based on the estimated noise variance $\widehat{\sigma_W^2}$ and the Eq. 8, we set the threshold as:

$$\lambda_D = \widehat{\sigma_W^2}\ (Q^{-1}(P_{fa})\sqrt{2N} + N)) \quad (29)$$

Where the $Q^{-1}(.)$ is the inverse of Q-function, $P_{fa}$ is the target probability of false alarm, and N is the number of samples.

### B. Experimental Setup

To evaluate the efficiency of the proposed model in a real-world scenario, we have implemented this model using GNU radio software and USRP units. Fig. 2 presents the experimental setup consisting of one transmitter and one receiver.

At the transmitter, we generated a signal of target SNR. This signal is transmitted using a USRP unit to another one. The two USRP units are connected through a radiofrequency cable to reduce the external noise. The received signal is used then to calculate the decision statistic, estimate the noise, and then get the sensing decision. The flow graph for the $QPSK$ signal generation was created using the blocks present in the GNU Radio software connected to the USRP transmitter by means of the $UHD\ USRP\ sink$ block. Gaussian noise was added to the signal source is varied by changing the standard deviation.

A second flow graph was created to get the samples from the USRP receiver and saved them in a Numpy array. The block $UHD\ source$ allows the GNU Radio software to obtain the samples from the USRP units. The FFT samples are squared and averaged over N samples to calculate the decision statistic $TED$. The noise variance is measured from the received signal based on the eigenvalues of the sample covariance matrix of the received signal $y[n]$, which is used to further compute the threshold $\lambda_D$, from Eq. 29. The decision statistic is then compared with the computed threshold to decide on the presence or the absence of the PU signal. If $TED \geq \lambda_D$ then the PU signal is present and if $TED \geq \lambda_D$ then the PU signal is absent.

## III. RESULTS

Using the experimental setup specified in the previous section, several experiments were performed to evaluate the efficiency of the proposed model. The performance of this technique with a dynamic threshold whose formula is given by Eq. 29 is evaluated and compared to that one of the energy detection with a static threshold using the probability of detection and the probability of false alarm. The static threshold is given by:

$$\lambda_s = F * (Q^{-1}(P_{fa})\sqrt{2N} + N) \quad (30)$$

Where $P_{fa}$ is the target probability of false alarm, $N$ is the number of samples, $Q^{-1}$ is the inverse of Q-function, and $F$ is the threshold factor.

The probability of detection and the probability of false alarm are computed as:

$$P_d = \frac{Number\ of\ detection}{number\ of\ trials} \quad (31)$$

$$P_{fa} = \frac{number\ of\ false\ alarm}{Number\ of\ trials} \quad (32)$$

To evaluate the impact of the threshold on the performance of energy detection, we fixed the number of samples to N=128, the probability of false alarm to 10%, and the static threshold to 148. This threshold is varied by multiplying this static threshold by the factor, 1, 1.5, 2, and 2.5 for SNR values ranging from $-20\ dB$ to $+20\ dB$. For each value of SNR, we calculated the probability of detection. Fig. 3 shows the probability of detection as a function of SNR for the aforementioned values of threshold factor. From this figure, it can be seen that the probability of detection increases as SNR increases. It can also be seen that the highest probability of detection corresponds to a value of threshold factor equal to 1. As we increase the value of the threshold factor, the probability of detection decreases.

To evaluate the impact of dynamic threshold selection on the probability of detection, we fixed the number of samples to N=128, the value of probability of false alarm to 10% and 20%, and we varied the value of SNR. For each value of SNR, we calculated the probability of detection of the sensing technique with a static and a dynamic threshold. Fig. 4 and Fig. 5 show the probability of detection as a function of SNR for the values of probability of false alarm of 10% and 20%, respectively. From these two figures, one can see that the probability of detection increases as SNR increases. For a value of probability of false alarm of 10% (Fig. 3) the probability of detection corresponding to the sensing technique with a dynamic threshold reaches 100% for a value of $SNR = -2\ dB$ while the one corresponding to the sensing technique with a static threshold reaches 100% for $SNR = +1\ dB$. For probability of false alarm of 20% (Fig. 4), the probability of detection corresponding to the sensing technique with a dynamic threshold reaches the value of 100% for a value of $SNR = -3\ dB$ while the one corresponding to the sensing technique with a static threshold reaches 100 % for a value of $SNR = 0\ dB$. As a first conclusion, the dynamic selection of the sensing threshold based on the noise estimation improves the detection performance by increasing the probability of detection.

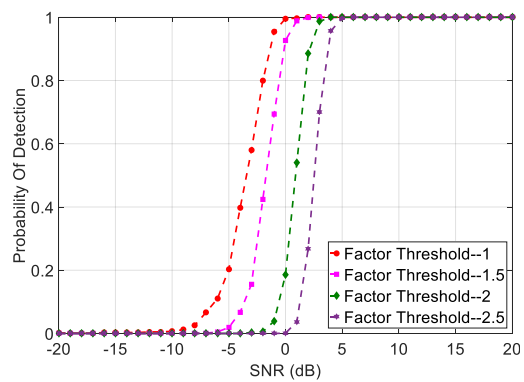

Fig. 3. Probability of detection versus probability of false alarm for snr=-5 dB and different values factor threshold

To evaluate the impact of the dynamic selection of the threshold on the probability of false alarm, we fixed the number of samples to 128, the value of SNR, and then we varied the probability of false alarm from 1% to 100%. For each value of the probability of false alarm, we calculated the probability of detection energy detection with a static and dynamic threshold.

Fig. 4. Probability of detection versus SNR for probability of false alarm 10%

Fig. 6. Probability of detection versus probability of false alarm for SNR= -20 dB and N=128

Fig. 5. Probability of detection versus SNR for probability of false alarm 20%

Fig. 7. Probability of detection versus probability of false alarm for SNR= -10 dB and N=128

Fig. 9. Probability of detection versus probability of false alarm for SNR= -2 dB and N=128

Fig. 8. Probability of detection versus probability of false alarm for SNR=-5 dB and N=128

Fig. 6 to Fig. 9 show the probability of detection as a function of the probability of False alarm for different values of SNR, $-20\ dB, -10\ dB, -5\ dB$, and $-2dB$, respectively. From these figures, it can be observed that the probability of detection increases at the probability of false alarm increases. Fig. 6 ($SNR = -20\ dB$) shows that the probability of detection with the dynamic threshold reaches the value of 100% for a value of probability of false alarm 62% while that one corresponding to the sensing technique with a static threshold reaches the value of 100% for a probability of false alarm 100%. As we increase the value of SNR, the probability of detection of both techniques reaches 100% faster. For instance, one can see from Fig. 9 ($SNR = -2dB$) that the probability of detection with dynamic threshold reaches 100% for a probability of false alarm of 10% while the one with static threshold reaches 100% for a probability of 90%. It can also be seen that the dynamic selection of the threshold increases the probability of detection and decreases the probability of false alarm. For instance, for a value of $SNR = -2dB$, if the target probability of detection is 100%, based on the result shown in Fig. 9, for the sensing technique with static threshold, we have to fix the probability of false alarm to a value higher than 90% while for sensing technique with a dynamic threshold, this probability of false alarm can be decreased to 10%. As a conclusion, the dynamic selection of the threshold increases the probability of detection and decreases the probability of false alarm.

IV. CONCLUSION

In this paper, we described an enhanced energy detection based technique to increase the probability of detection and decrease the probability of false alarm using a dynamic threshold selection based on measuring the noise level present in the received signal. The level of noise is measured using a blind technique based on sample covariance matrix eigenvalues of the received signal. The proposed approach was implemented using the GNU Radio software and USRP units. The results show that the dynamic selection of the sensing threshold proposed in this work increases the probability of detection and decreases the probability of false alarm.

ACKNOWLEDGMENT

The authors acknowledge the support of the US Fulbright program.

REFERENCES

[1] N. Kaabouch and W. C. Hu, Handbook of research on software-defined and cognitive radio technologies for dynamic spectrum management. IGI Global, 2014.

[2] A. Ali and W. Hamouda, "Advances on Spectrum Sensing for Cognitive Radio Networks: Theory and Applications," *IEEE Commun. Surv. Tutorials*, pp. 1–1, 2016.

[3] Y. Zeng, Y.-C. Liang, A. T. Hoang, and R. Zhang, "A Review on Spectrum Sensing for Cognitive Radio: Challenges and Solutions," *EURASIP J. Adv. Signal Process.*, vol. 2010, pp. 1–16, 2010.

[4] A. Ranjan, Anurag, and B. Singh, "Design and analysis of spectrum sensing in cognitive radio based on energy detection," *International Conference on Signal and Information Processing*, 2016, pp. 1–5.

[5] R. T. Khan, M. I. Islam, S. Zaman, and M. R. Amin, "Comparison of cyclostationary and energy detection in cognitive radio network," *International Workshop on Computational Intelligence*, 2016, pp. 165–168.

[6] M. R. Manesh, M. S. Apu, N. Kaabouch, and W.-C. Hu, "Performance evaluation of spectrum sensing techniques for cognitive radio systems," *IEEE Ubiquitous Computing, Electronics & Mobile Communication Conference*, 2016, pp. 1–7.

[7] R. Tandra and A. Sahai, "SNR Walls for Signal Detection," *IEEE J. Sel. Top. Signal Process.*, vol. 2, no. 1, 2008.

[8] N. Giweli, S. Shahrestani, and H. Cheung, "selection of spectrum sensing method to enhance qos in cognitive radio networks," *Int. J. Wirel. Mob. Networks*, vol. 8, no. 1, 2016.

[9] H. Reyes, S. Subramaniam, N. Kaabouch, and W. C. Hu, "A spectrum sensing technique based on autocorrelation and Euclidean distance and its comparison with energy detection for cognitive radio networks," *Comput. Electr. Eng.*, vol. 52, no. C, pp. 319–327, May 2016.

[10] S. Subramaniam, H. Reyes, and N. Kaabouch, "Spectrum occupancy measurement: An autocorrelation based scanning technique using USRP," *IEEE Wireless and Microwave Technology Conference*, 2015, pp. 1–5.

[11] M. R. Manesh, S. Subramania, H. Reyes, and N. Kaabouch, "Real-time Spectrum Occupancy Monitoring Using a Probabilistic Model," *Computer Networks, Elsevier*, 2017.

[12] H. Reyes, S. Subramanian, N. Kaabouch, and W. C. Hu, "A Bayesian Inference Method for Scanning the Radio Spectrum and Estimating the Channel Occupancy," *IEEE Annual Ubiquitous Computing, Electronics & Mobile Communication Conference*, pp. 1-6, 2016.

[13] F. Salahdine, H. El Ghazi, N. Kaabouch, and W. F. Fihri, "Matched filter detection with dynamic threshold for cognitive radio networks," *International Conference on Wireless Networks and Mobile Communications*, 2015, pp. 1–6.

[14] X. Zhang, R. Chai, and F. Gao, "Matched filter based spectrum sensing and power level detection for cognitive radio network," *IEEE Global Conference on Signal and Information Processing*, 2014, pp. 1267–1270.

[15] Y. Arjoune and N. Kaabouch "Wideband Spectrum Scanning: an approach based on Bayesian Compressive Sensing" *Jounal of IET communications*, Accepted, 2017.

[16] Y. Arjoune, N. Kaabouch, H. El Ghazi, and A. Tamtaoui, "Compressive Sensing: Performance Comparison Of Sparse Recovery Algorithms " *The IEEE Annual Computing and Communication Workshop and Conference*, pp. 1-6, 2017.

[17] D. R. Joshi, D. C. Popescu, and O. A. Dobre, "Adaptive spectrum sensing with noise variance estimation for dynamic cognitive radio systems," *Conference on Information Sciences and Systems (CISS)*, 2010, pp. 1–5.

[18] A. Muralidharan, P. Venkateswaran, S. G. Ajay, D. Arun Prakash, M. Arora, and S. Kirthiga, "An adaptive threshold method for energy based spectrum sensing in Cognitive Radio Networks," *International Conference on Control, Instrumentation, Communication and Computational Technologies*, 2015, pp. 8–11.

[19] M. Sarker, "Energy detector based spectrum sensing by adaptive threshold for low SNR in CR networks," *Wireless and Optical Communication Conference*, 2015, pp. 118–122.

[20] J. Wu, T. Luo, and G. Yue, "An Energy Detection Algorithm Based on Double-Threshold in Cognitive Radio Systems," *International Conference on Information Science and Engineering*, 2009, pp. 493–496.

[21] S. Suwanboriboon and W. Lee, "A novel two-stage spectrum sensing for cognitive radio system," *International Symposium on Communications and Information Technologies (ISCIT)*, 2013, pp. 176–181.

[22] D. M. M. Plata and Á. G. A. Reátiga, "Evaluation of energy detection for spectrum sensing based on the dynamic selection of detection-threshold," *Procedia Eng.*, vol. 35, pp. 135–143, 2012.

[23] J. Zhu, Z. Xu, F. Wang, B. Huang, and B. Zhang, "Double Threshold Energy Detection of Cooperative Spectrum Sensing in Cognitive Radio," *International Conference on Cognitive Radio Oriented Wireless Networks and Communications*, 2008, pp. 1–5.

[24] M. Hamid, N. Bjorsell, and S. Ben Slimane, "Sample covariance matrix eigenvalues based blind SNR estimation," *IEEE International Instrumentation and Measurement Technology Conference Proceedings*, 2014, pp. 718–722.

[25] M. R. Manesh, A. Quadri, S. Subramaniam, and N. Kaabouch, "An optimized SNR estimation technique using particle swarm optimization algorithm," *IEEE Computing and Communication Workshop and Conference*, 2017, pp. 1–6.

[26] M. R. Manesh, N. Kaabouch, and H. Reyes, "A Bayesian Approach to Estimate and Model SINR in Wireless Networks," *International Journal of Communication Systems, Wiley*, pp. 1-11, August 2016.

[27] A. Quadri, M. R. Manesh, and N. Kaabouch, "Performance Comparison of Evolutionary Algorithms for Noise Cancellation in Cognitive Radio Systems," *The IEEE Annual Computing and Communication Workshop and Conference*, pp. 1-6, 2017.